\pdfoutput=1
\documentclass[10pt,conference,letterpaper]{IEEEtran}
\usepackage{amsmath}
\usepackage{graphicx}
\usepackage{cite}
\usepackage{color}
\usepackage{url}
\newtheorem{proposition}{\bf   Proposition}
\newtheorem{theorem}{\bf Theorem}

\IEEEoverridecommandlockouts

\begin{document}

%\sloppy

\title{
The Impact of Small-Cell Bandwidth Requirements on Strategic Operators \thanks{This work was
supported by NSF under grant AST-134338.}
%A Pricing Model for Optimizing Femto-Cell Deployments With Additional Unlicensed Access
%\thanks{This work was supported by NSF under grant 1343381.}
}

\author{
\IEEEauthorblockN{Cheng Chen}
\IEEEauthorblockA{NGS, Intel Corporation \\
Hillsboro, OR 97124\\
cheng.chen@intel.com}
\and
\IEEEauthorblockN{Randall A. Berry, Michael L. Honig}
\IEEEauthorblockA{Dept. of EECS, Northwestern University \\
Evanston, IL 60208\\
\{rberry, mh\}@eecs.northwestern.edu}
\and
\IEEEauthorblockN{Vijay G. Subramanian}
\IEEEauthorblockA{Dept. of EECS, University of Michigan\\
Ann Arbor, MI 48109\\
vgsubram@umich.edu}
}
\maketitle

\begin{abstract}
Small-cell deployment in licensed and unlicensed spectrum is considered to be one
of the key approaches to cope with the ongoing wireless data demand explosion. Compared to traditional
cellular base stations with large transmission power, small-cells typically have
relatively low transmission power, which makes them attractive for some spectrum
bands that have strict power regulations, for example, the 3.5GHz band \cite{FCC}. In
this paper we consider a heterogeneous wireless network consisting of one or more
service providers (SPs). Each SP operates in both macro-cells and small-cells, and provides
service to two types of users: mobile and fixed. Mobile users can only associate with
macro-cells whereas fixed users can connect to either macro- or small-cells. The SP charges a
price per unit rate for each type of service. Each SP is given
a fixed amount of bandwidth and splits it between macro- and small-cells. Motivated by bandwidth regulations, such as those for the 3.5Gz band, we assume a minimum amount of
bandwidth has to be set aside for small-cells.  We study the optimal pricing and bandwidth allocation strategies in both monopoly and competitive scenarios. In the monopoly scenario the strategy is unique. In
the competitive scenario there exists a unique Nash equilibrium, which depends on the regulatory constraints. We also analyze the social welfare achieved,
and compare it to that without the small-cell bandwidth constraints. Finally, we discuss implications of our results on the effectiveness of the minimum bandwidth constraint on influencing small-cell deployments.
%%%Short version
%Small cells deployed in licensed spectrum and unlicensed access
%via WiFi provide different ways of expanding wireless services
%to low mobility users. That reduces the demand for conventional
%macro-cellular networks, which are better suited for wide-area
%mobile coverage. The mix of these technologies seen in practice
%depends in part on the decisions made by wireless service providers
%that seek to maximize revenue, and allocations of licensed and
%unlicensed spectrum by regulators. To understand these interactions
%we present a model in which a service provider allocates available
%licensed spectrum across two separate bands, one for macro- and
%one for small-cells, in order to serve two types of users: mobile and fixed.
%We assume a service model in which the providers can charge
%a (different) price per unit rate for each type of service
%(macro- or small-cell); unlicensed access is free.
%With this setup we study how the addition of unlicensed spectrum
%affects prices and the optimal allocation of bandwidth
%across macro-/small-cells. We also characterize the optimal fraction
%of unlicensed spectrum when new bandwidth becomes available.

\end{abstract}

\section{Introduction}
Current cellular networks are expected to evolve towards heterogeneous networks (HetNets) to cope with the explosive demand for wireless data \cite{HetNet1-Qualcomm11,HetNet2-Ghosh12}. This requires service providers (SPs) to deploy small-cells in addition to traditional macro-cells. While typical macro-cells, such as cellular base stations, typically have large transmission power and therefore are capable of covering users within a large region, small-cells have much lower transmission power and are  used to provide service to a local area. This unique characteristic of small-cells enables them to be an attractive choice in some spectrum bands that have strict power regulations. For example, in 2012, the FCC proposed to create a new Citizens Broadband Service in the 3550-3650 MHz band (3.5GHz Band), previously utilized for military and satellite operations \cite{FCC}. Due to the low power constraint within this band, only small-cells can be deployed. For SPs that want to use this band to expand their service, this type of bandwidth regulation needs to be taken into account in determining optimal resource allocation strategies.

While the deployment of small-cells will increase overall data capacity, it also complicates the network management and resource allocation for SPs. This includes how to differentiate the pricing schemes and optimally split their limited bandwidth resources between macro- and small-cells, taking into account the fact that users in the network are also heterogeneous in terms of mobility patterns. Moreover, these decisions are further complicated by regulatory restrictions on certain bands, such as the designation of new spectrum in the 3.5GHz band only for small-cells.

\subsection{Contributions}

Our paper analyzes the impact of regulatory requirements that certain bands be used only for small-cells on competitive service providers that allocate bandwidth between macro- and small-cell networks. We also analyze the associated social welfare. At present new spectrum is typically apportioned based on an auction, and another goal is to provide insight into the social welfare achieved via winner-take-all auctions. Given the policy implications of such an analysis, we briefly discuss these results at a high-level; detailed results are in Section~\ref{Sec:SW}.

The scenario that we consider in the paper is the following. The spectrum regulator needs to allocate $B$  units of newly available bandwidth to two competitive SPs. Each SP has an initial endowment of licensed bandwidth $B_{1}^o$ and $B_{2}^o$, and gets a proportion of the new bandwidth, denoted as $B_{1}^n$ and $B_2^n$. The regulator determines the rules for this assignment using an appropriate auction procedure; e.g., an allocation of $B_{1}^n=B$ and $B_{2}^n=0$ corresponds to the outcome of a winner-take-all auction that SP 1 wins.The initial bandwidth can be allocated by each SP to either macro-cells or small-cells. In contrast, the new bandwidth can only be used for small-cells, as enforced by the regulatory constraint.

\begin{figure}[htbp]
\centering
\includegraphics[width=0.45\textwidth,height=0.35\textwidth]{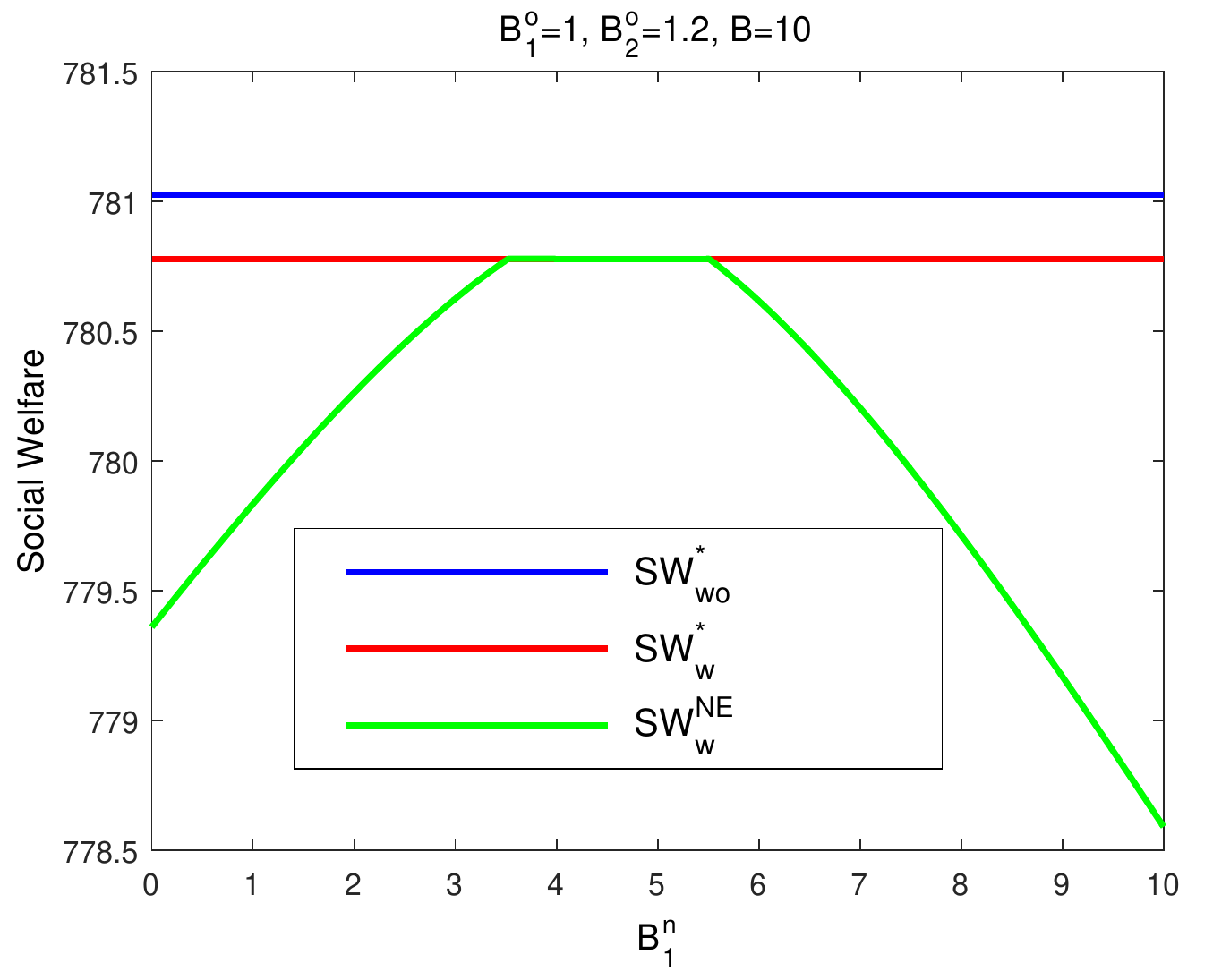}
\caption{Social welfare versus $B_1^n$ with large $B$. Here $\text{SW}_{\text{wo}}^{*}$ is the optimal social welfare without regulatory constraints, $\text{SW}_{\text{w}}^{*}$ is the optimal social welfare with regulatory constraints, and $\text{SW}_{\text{w}}^{\text{NE}}$ is the equilibrium social welfare with regulatory constraints.}
\label{Fig:SW_2}
\end{figure}

In Figure~\ref{Fig:SW_2} we present a typical result that we obtain for different partitions of the new bandwidths amongst the two SPs. The blue line represents the social welfare achieved when the SPs cooperate and the new bandwidth comes with no restrictions; the red line represents the social welfare achieved when the SPs cooperate and the new bandwidth is restricted to small cell use; and the green curve represents the social welfare achieved when the SPs compete with the new bandwidth restricted only to small cell use. From the results of our previous work \cite{Competition5-Chen15}, it is easily verified that blue line is also the social welfare achieved when the SPs compete and there is no restriction on the usage of the new bandwidth. It is clear from Figure~\ref{Fig:SW_2} that the introduction of restrictions on the usage of new bandwidth results in a reduction in social welfare, and additionally, only specific partitions of the new bandwidth will lead to this reduced social welfare value being achieved with competing SPs. It should also be noted that the assignment that results from a winner-take-all auction yields much lower social welfare; these correspond to the two endpoints in the figure. Numerical investigations also show that the SP with the larger amount of initial bandwidth endowment obtains the highest marginal revenue increase from any new bandwidth with the other SP losing revenue when this occurs. The larger SP would thus, bid higher to reduce the influence of the smaller SP.

Thus one of the main contributions of our paper is to highlight the possibility of such negative outcomes with the specific designation chosen for small cells, and also to point out the necessity of carrying out such analysis before deciding on other regulatory constraints for newly available spectrum bands.

%We now summarize our main contributions in this paper:
We now summarize our other contributions in this paper:

1. \emph{Incorporating Bandwidth Regulations into the HetNet Model}: Prior related work that considers bandwidth allocation assumes SPs are free to split spectrum between macro- and small-cells in any way. Here,  we add additional small-cell bandwidth constraints that impose a minimum amount of small-cell bandwidth to the SPs. This is primarily motivated by the 3.5GHz Band released by the FCC that can only be used to deploy small-cells. The introduction of such bandwidth constraints has a direct influence on the optimal pricing and bandwidth allocation strategies of SPs.

2. \emph{Characterizing the impact of the regulatory constraints for SPs in both monopoly and competitive scenarios}: We analyze scenarios with both a monopoly SP and competitive SPs. We show that in the monopoly scenario the SP simply increases its small-cell bandwidth to the required minimum amount if its original small-cell bandwidth is less than the constraint. This applies to both social welfare and revenue-maximization. With two competitive SPs, there always exists a unique Nash equilibrium that depends on the regulatory constraints.  We illustrate this by considering three cases corresponding to whether the original equilibrium allocation satisfies the two constraints. We characterize the equilibrium for each case. %We also use a specific example to illustrate the different parameter regions corresponding to different types of equilibrium.

3. \emph{Social Welfare Analysis}: We also quantify the influence of the regulatory constraints on the social welfare. We conclude that if the equilibrium without constraints violates the constraints, then social welfare loss is inevitable. However, the social welfare loss is always bounded, and the worst case happens when the spectrum regulator requires the SPs to allocate all bandwidth only to small-cells. In this extreme case there are no macro-cells, and consequently none of the mobile users receives wireless service.

%4. \emph{Thoughts on the effects of regulatory constraints}: For a spectrum regulator, such as the FCC, a common challenge is to determine the optimal bandwidth split between different SPs once some new spectrum becomes available. We develop a simple model to analyze this case for the 3.5GHz Band. Specifically, we assume the regulator needs to split the new bandwidth between two competitive SPs, where the new bandwidth can only be allocated to small-cells. We give the conditions for the existence of optimal split schemes that can achieve the optimal social welfare without the impact of such regulatory constraints. If the conditions are satisfied, we also present the optimal bandwidth split schemes. However, we also notice that if the amount of new bandwidth is large compared to the original bandwidth owned by the SPs, social welfare loss is inevitable under any bandwidth split schemes. This observation suggests that the spectrum regulator needs to be careful when designing the regulatory constraints for newly available spectrum bands, and the regulations should not change the original market equilibrium too much. Otherwise social welfare loss would be incurred.

\subsection{Related work}

Pricing and bandwidth allocation problems in HetNets have attracted considerable attention. In \cite{Opt1-Shetty09, Opt2-Gussen11, Opt3-Yun12}, small-cell service is seen as an enhancement to macro-cell service. In contrast, small-cell and macro-cell service are considered to be separate services in \cite{Opt4-Chen11,Opt5-Lin11,Opt6-Duan13,Opt7-Chen13,Competition1-Zhang13, Competition2-Hossain08, Competition5-Chen15,Investment-Chen16, Unlicensed1-Chen16}, the same as our model in this paper. Only optimal pricing is studied in \cite{Opt1-Shetty09, Opt3-Yun12, Competition1-Zhang13, Competition2-Hossain08}, while \cite{Opt2-Gussen11, Opt4-Chen11,Opt5-Lin11,Opt6-Duan13,Opt7-Chen13,Competition5-Chen15,Investment-Chen16,Unlicensed1-Chen16} consider joint pricing and bandwidth allocation, as in this paper. Additionally, except for \cite{Competition1-Zhang13, Competition2-Hossain08,Competition5-Chen15,Unlicensed1-Chen16, Investment-Chen16} that include the competitive scenario with multiple SPs, all the other work assumes only one SP. In this paper, we  investigate both monopoly and competitive scenarios, and adopt a model similar to that in our previous work \cite{Opt7-Chen13, Competition5-Chen15} (which did not consider bandwidth regulations).

%We first study the monopoly scenario and characterize the impacts. We then switch to the competitive scenario with two SPs; extending to beyond two does not change the story qualitatively but merely increases the complexity of analysis owing to the many combinatorial possibilities of the specific constraints being tight or not. We find that although the Nash equilibrium is still unique, it can be of different types depending on the specific values assumed for the regulatory constraints. Furthermore, the added regulatory constraint can also incur a social welfare loss compared to the case without the constraints. Finally, using the analysis developed we address the general scenario where the spectrum regulator has additional newly available bandwidth that can only be used in small-cells and wants to determine the optimal bandwidth assignment strategy between two competing SPs for the spectrum regulator in terms of maximizing social welfare. Having characterized this we make high-level recommendations on how this assignment process should be designed.

The rest of the paper is organized as follows. We present the system model in Section \ref{Sec:System Model}. We consider monopoly and competitive scenarios in Section \ref{Sec:Monopoly} and Section \ref{Sec:Competitive}, respectively. Social welfare analysis is in Section \ref{Sec:SW}. We conclude in Section \ref{Sec:Conclusions}. All proofs of the main results can be found in the appendices.

\section{System Model}\label{Sec:System Model}
We adopt the mathematical model in our previous work \cite{Competition5-Chen15} for the analysis. We now describe the different aspects of it while pointing out the additional elements considered here.

\subsection{SPs}
We consider a HetNet with $N$ SPs providing separate macro- and small-cell service to all users. Denote the set of SPs as $\mathcal{N}$. Each SP is assumed to operate a two-tier cellular network consisting of macro- and small-cells that are uniformly deployed over a given area. We further assume all SPs have the same density of infrastructure deployment. We normalize the density of macro-cells to one. The density of small-cells is denoted as $N_S$. In our setting, macro-cells have high transmission power, and therefore can provide large coverage range. In contrast, small-cells have low transmission power, and consequently local coverage range.

Each SP $i$ has a total amount of bandwidth $B_i$ exclusively licensed.\footnote{For the monopoly SP scenario, we will ignore the subscript.} Since we assume all macro- and small-cells use separate bands, each SP $i$ needs to decide how to split its bandwidth into $B_{i,M}$, bandwidth allocated to macro-cells, and $B_{i,S}$, bandwidth allocated to small-cells. When determining this partition, every SP is required to conform to (possible) bandwidth regulations enforced by the spectrum regulator. Specifically, SP $i$ is requested to guarantee a minimum amount of bandwidth allocated to small-cells, and this lower bound is denoted as $B_{i,S}^0$.

For a fixed bandwidth allocation, the total achievable data rate provided by the macro-cells of SP $i$ is $C_{i, M}=B_{i, M}R_0$, where $R_0$ is the (average) spectral efficiency of the macro-cells. The total available rate in small-cells of SP $i$ is given by $C_{i, S}=\lambda_S B_{i, S}R_0$, where $\lambda_S>1$ reflects the increase in spectral efficiency due to smaller cell size, and possibly greater deployment density. Each SP $i$ provides separate macro- and small-cell services and charges the users a price \emph{per unit rate} for associating with its macro-cells or small-cells, namely, $p_{i, M}$ and $p_{i, S}$.

\subsection{Users}
We assume the users in the networks are also heterogeneous and categorize them into two types based on their mobility patterns. Mobile users can only be served by macro-cells. In contrast, fixed users are relatively stationary, and can connect to either macro- or small-cells (but not both). Denote the densities of mobile users and fixed users as $N_m$ and $N_f$, respectively. Note that the heterogeneity of the users can also arise from an equivalent model that assumes $(N_m+N_f)$ as the total density of users, who are mobile with probability $N_m/(N_m+N_f)$ and stationary with probability $N_f/(N_m+N_f)$. After user association, let $K_{i, M}$ and $K_{i, S}$ denote the mass of users connected to the macro- and small-cells of SP $i$, respectively. (Note that $K_{i, S}$ consists of fixed users only, whereas $K_{i, M}$ can consist of both mobile and fixed users.)

\begin{figure}[htbp]
\centering
\includegraphics[width=0.45\textwidth,height=0.3\textwidth]{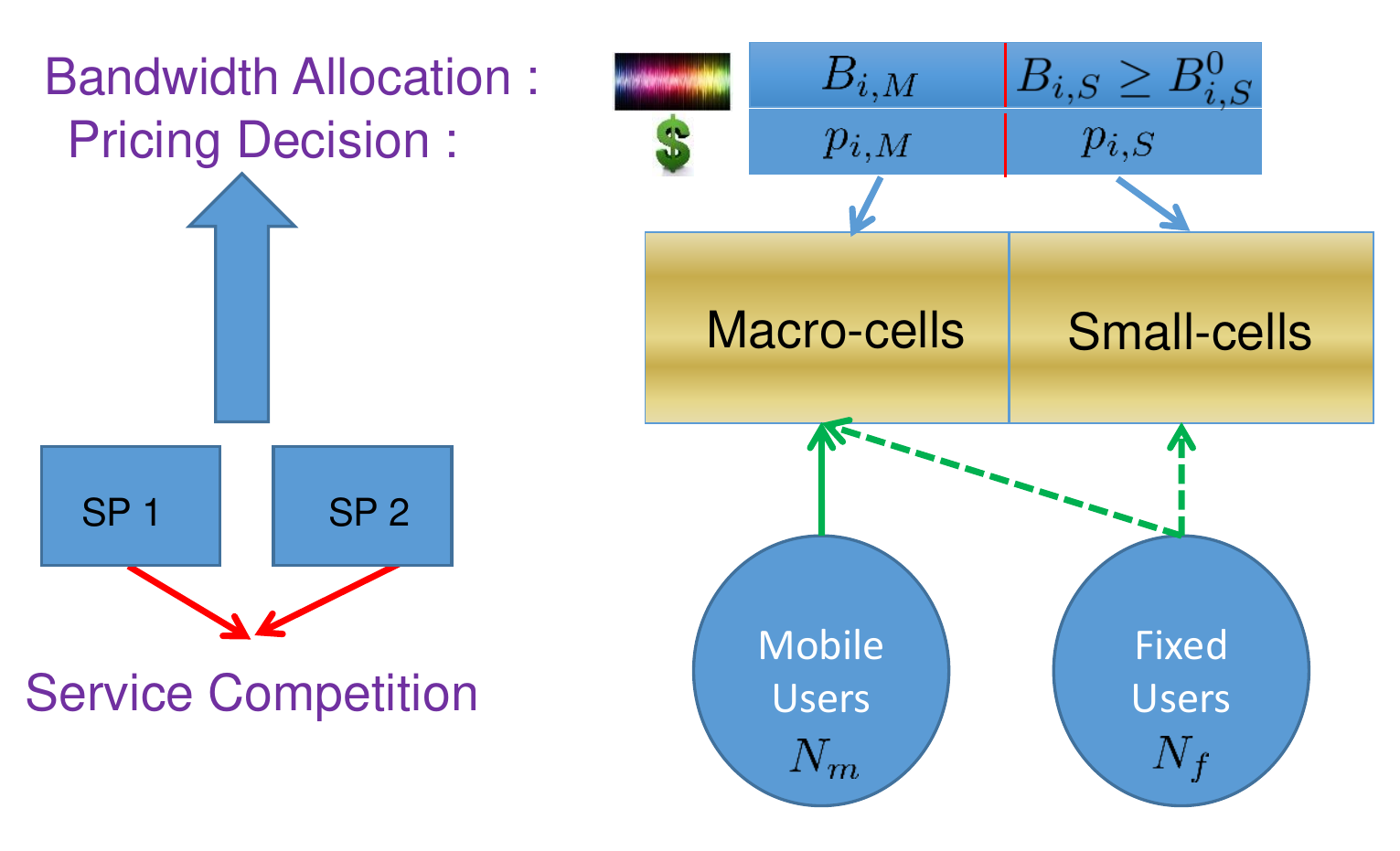}
\caption{System Model.}
\label{Fig:System_Model}
\end{figure}

\subsection{User and SP Optimization}
Figure \ref{Fig:System_Model} illustrates the network and market model. We now introduce the optimization problems corresponding to both users and SPs. We assume each user is endowed with a utility function, $u(r)$, which only depends on the service rate it gets. For simplicity of analysis, in this paper we assume that all users have the same $\alpha$-fair utility functions \cite{MoWalrand} with $\alpha \in (0,1)$:
\begin{equation}\label{Eqn:UtilityFn}
u(r)=\frac{r^{1-\alpha}}{1-\alpha}, \quad \alpha \in (0,1).
\end{equation}
This restriction enables us to explicitly calculate many equilibrium quantities, which appears to be difficult for more general classes of utility. Furthermore, this class is widely used in both networking and economics, where it is a subset of the class of iso-elastic utility functions.\footnote{In general $\alpha$-fair utilities require that $\alpha \geq 0$ to ensure concavity; requiring $\alpha>0$ ensures strict concavity but allows us to approach the linear case as $\alpha \rightarrow 0$.  The restriction of $\alpha<1$  ensures that utility is non-negative so that a user can always ``opt out" and receive zero utility. Note also that as $\alpha \rightarrow 1$, we approach the  $\log(\cdot)$ (proportional fair) utility function.}

Each user chooses the service by maximizing its net payoff $W$,
defined as its utility less the service cost. For a service with price $p$, this is equivalent to:
\begin{equation} \label{Opt:User Optimization}
W= \max_{r\geq 0}\quad u(r)-pr.
\end{equation}
For $\alpha$-fair utility functions,  (\ref{Opt:User Optimization}) has the  unique solution:
\begin{equation}\label{Eqn:UserRateOpt}
r^{*}= D(p)=(u^{\prime})^{-1}(p)=(1/p)^{1/\alpha}, %\frac{1}{\sqrt[\alpha]{p}}%\max \Big((u^\prime)^{-1}(p),0 \Big)
\end{equation}
where $D(p)$ here can be seen as the user's rate demand function. The maximum net payoff for a user is thus:
\begin{equation} \label{Eqn:User Net Payoff}
W^*(p) = u(D(p))- pD(p)=\frac{\alpha}{1-\alpha}p^{1-\frac{1}{\alpha}}.
\end{equation}
Recall that fixed users can choose between any macro- or small-cell service offered by any SP, while mobile users can only choose the macro-cell service provided by a SP. However, here, we assume mobile users have priority connecting to macro-cells, which means macro-cells will only admit fixed users after the service requests of all mobile users have been addressed.

For the association rules, we adopt the same process described in \cite{Competition5-Chen15}. That is, users always choose the service with lowest price and fill the corresponding capacity. If multiple services have the same price, then the users are allocated across them in proportion
to the capacities. Once a particular service capacity is exhausted, then the leftover demand continues to fill the remaining service in the same fashion.

Each SP determines the bandwidth split and service prices to maximize its revenue, which is the aggregate amount paid by all users associating with their macro- and small-cells. Meanwhile they also need to conform to the constraints on small-cell bandwidth allocation. Specifically, SP $i$ solves the following optimization problem:
\begin{subequations}
\begin{align}
\text{maximize}  \quad &S_i=p_{i, M}K_{i, M}D(p_{i, M})+p_{i, S}K_{i, S}D(p_{i, S}), \label {Opt:SP's optimization} \\
\text{subject to}  \quad& B_{i, M}+B_{i, S}\le B_i,  B_{i, M}\ge 0, B_{i, S}\ge B_{i,S}^0,  \label {Opt:SP's optimization-Constraint1}\\
 & 0<p_{i, M}, p_{i, S}<\infty. \label {Opt:SP's optimization-Constraint2}
\end{align}
\end{subequations}

Alternatively, a social planner, such as the FCC, may seek to allocate bandwidth and set prices
to maximize social welfare, which is the sum utility of all users, subject to the same constraints (\ref{Opt:SP's optimization-Constraint1}) and (\ref{Opt:SP's optimization-Constraint2}).
This is given by:
\begin{align}
\text{maximize}  \quad &\text{SW}=\sum\limits_{i=1}^{N}[K_{i, M}u(D(p_{i, M}))+K_{i, S}u(D(p_{i, S}))].\label{Opt:SW optimization}
\end{align}

\subsection{Sequential Game and Backward Analysis}
We model the bandwidth and price adjustments of SPs in the network as a two-stage process:
\begin{enumerate}
\item
Each SP $i$ first determines its bandwidth allocation $B_{i, M}, B_{i, S}$ between macro-cells and small-cells. Denote the aggregate bandwidth allocation profile as $\mathbf{B}$.
\item
Given $\mathbf{B}$ (assumed known to all SPs), the SPs announce prices for both macro-cells and small-cells. The users then associate with SPs according to the previous user association rule.
\end{enumerate}

%This order reflects the fact that bandwidth partitioning takes place over a slower time-scale than price adjustments, since changing the bandwidth partition could conceivably involve reconfiguring equipments at both base stations and handsets, and adjusting the placement of access points along with transmission parameters in order to keep the rate per cell fixed. Adjustment of prices would not require these additional changes. We implicitly assume, however, that price adjustments happen slow enough to allow the users to re-associate according to the previous association rules.

We then do backward induction. That is, we first derive the price equilibrium under a fixed bandwidth allocation. We then characterize the bandwidth allocation equilibrium based on the price equilibrium obtained.
%In the monopoly scenario, the two-stage process can be seen as a sequential optimization procedure. That is, the single SP first determines optimal pricing with fixed bandwidth allocation, and then optimizes the bandwidth allocation based on the optimal pricing scheme with the bandwidth allocation fixed.
%
%If there are multiple SPs, the SPs compete for a finite population of mobile and fixed users, and the two-stage process is thus a sequential game. To gain insight into the resulting allocations we therefore seek a subgame-perfect Nash equilibrium consisting of the following:
%({\it i}) A price equilibrium based on each fixed bandwidth allocation;
%and ({\it ii}) A bandwidth allocation equilibrium given that prices are set according to ({\it i}).

\section{Monopoly Scenario with A Single SP}\label{Sec:Monopoly}
We first study the bandwidth allocation when a single SP is operating in the network. This is similar to the analysis in our previous work \cite{Opt7-Chen13}, except here we have an additional regulatory constraint that imposes a minimum bandwidth allocation to small-cells. This added constraint will change the optimal bandwidth allocation strategy for the monopoly SP. In \cite{Opt7-Chen13} it is concluded that for the set of $\alpha$-fair utility functions we use in this paper, the revenue-maximizing and social welfare-maximizing bandwidth allocation turn out to be the same. The following theorem states that the optimal bandwidth allocations under both objectives are still the same, but adding a large value for the bandwidth set aside for small-cells changes the optimal bandwidth allocation.

\begin{theorem}[Optimal Monopoly Bandwidth Allocation] \label{Thm:Monopoly Bandwidth Allocation}
For a monopoly SP, the optimal revenue-maximizing and social welfare-maximizing bandwidth allocation strategies are the same and can be determined by the following cases:

1. If $B_S^0\le \frac{N_f\lambda_S^{1/\alpha-1}B}{N_f\lambda_S^{1/\alpha-1}+N_m}$, the optimal bandwidth allocation remains the same as that without the regulatory constraint. In this case it is given by:
\begin{subequations}
\begin{align}
&B_S^{\text{SW}}=B_S^{\text{rev}}=\frac{N_f\lambda_S^{1/\alpha-1}B}{N_f\lambda_S^{1/\alpha-1}+N_m}, \\ &B_M^{\text{SW}}=B_M^{\text{rev}}=\frac{N_m B}{N_f\lambda_S^{1/\alpha-1}+N_m}.
\end{align}
\end{subequations}

2. If $B_S^0> \frac{N_f\lambda_S^{1/\alpha-1}B}{N_f\lambda_S^{1/\alpha-1}+N_m}$,  the optimal bandwidth allocation is changed to:
\begin{equation}
B_S^{\text{SW}}=B_S^{\text{rev}}=B_S^0, \quad B_M^{\text{SW}}=B_M^{\text{rev}}=B-B_S^0.
\end{equation}
Consequently there will be both a welfare and revenue loss if this case applies.

In both cases the optimal macro- and small-service prices are market-clearing prices, i.e., the prices that equalize the total rate demand and the total rate supply in both cells.
\end{theorem}

Theorem \ref{Thm:Monopoly Bandwidth Allocation} states that if the original optimal bandwidth allocation without the bandwidth regulations already satisfies the imposed constraint, then the SP just keeps the same bandwidth allocation. If the original bandwidth allocation violates the regulatory constraint, then the SP increases the small-cell bandwidth to the required level. This is because the added regulatory constraint does not change the concavity of the revenue or social welfare function with respect to the small-cell bandwidth, and further increasing the bandwidth allocation to small-cells will only lead to more revenue or social welfare loss.

\section{Competitive Scenario with Two SPs}\label{Sec:Competitive}
In this section we turn to the competitive scenario with two SPs, each of which maximizes its individual revenue. Applying the results from \cite{Competition5-Chen15}, the price equilibrium given any fixed bandwidth allocation is always the market-clearing price. We therefore focus on the bandwidth allocation Nash equilibrium.

Considering the case without the additional regulatory constraint, using the results from \cite{Competition5-Chen15}, there exists a unique Nash equilibrium and the bandwidth allocations of two SPs at equilibrium are given by:
\begin{subequations}
\begin{align}
&B_{1,S}^{\text{NE}}=\frac{N_f\lambda_S^{1/\alpha-1}B_1}{N_f\lambda_S^{1/\alpha-1}+N_m}, B_{1,M}^{\text{NE}}=\frac{N_m B_1}{N_f\lambda_S^{1/\alpha-1}+N_m},\\
&B_{2,S}^{\text{NE}}=\frac{N_f\lambda_S^{1/\alpha-1}B_2}{N_f\lambda_S^{1/\alpha-1}+N_m},  B_{2,M}^{\text{NE}}=\frac{N_m B_2}{N_f\lambda_S^{1/\alpha-1}+N_m}.
\end{align}
\end{subequations}

With the additional regulatory constraints, we have the following theorem characterizing the corresponding Nash equilibrium between two SPs.

\begin{theorem} \label{Thm:NE}
With two SPs, with a constraint on minimum small-cell bandwidth, the Nash equilibrium exists and is unique. Moreover, the total bandwidth allocated to small-cells by the two SPs is no less than that without the regulatory constraints.
\end{theorem}

Theorem \ref{Thm:NE} states that the existence and uniqueness of the Nash equilibrium is preserved after adding the regulatory constraints. This can be proved using similar methods as provided in our previous work \cite{Competition5-Chen15}, with some  modifications. The last part of the theorem may be more subtle than it appears. One may try to argue that if any of the constraints is violated, that SP then needs to increase its bandwidth allocation to small-cells. It would then hold that the total bandwidth allocated to small-cells surely increases. However, the logic does not carry through if only one constraint is violated at the Nash equilibrium omitting the constraint. In that case, the SP with violated constraint must increase the bandwidth allocation to small-cells. However, the other SP, whose equilibrium small-cell bandwidth allocation without regulations satisfies the constraint, may potentially \emph{decrease} its bandwidth in small-cells in response to the increase in bandwidth allocation of its competitor. In that case, determining the change in total bandwidth requires a more detailed analysis. Nonetheless, Theorem \ref{Thm:NE} indicates even in that case the total bandwidth in small-cells would not decrease. We will present a specific example later.

Depending on whether the regulatory constraints are violated or not at the Nash equilibrium without the constraints, there are three cases we need to cover independently. We will see that, in each case, the Nash equilibrium behaves differently.

{\it Case A: Both constraints are satisfied.}
The new Nash equilibrium is the same as the Nash equilibrium without regulations.

{\it Case B: Both constraints are violated.}
The Nash equilibrium without regulations is no longer valid. The following proposition characterizes the properties of the new Nash equilibrium.
\begin{proposition}\label{Prop:Both Violated NE}
In case B, the Nash equilibrium with regulatory constraints is one of the following types:\\
Type I. Both SPs increase their small-cell bandwidth allocations to exactly the required amount, i.e, $B_{1,S}=B_{1,S}^0, B_{2,S}=B_{2,S}^0$.\\
Type II. One SP increases its small-cell bandwidth exactly to the required amount, while the other SP increases further beyond the required amount, i.e, $B_{1,S}=B_{1,S}^0, B_{2,S}>B_{2,S}^0 \text{ or } B_{1,S}>B_{1,S}^0, B_{2,S}=B_{2,S}^0$.
\end{proposition}

It is conceptually easy to characterize the necessary and sufficient conditions for the first type of Nash equilibrium to hold since at that equilibrium the marginal revenue increase with respect to per unit of bandwidth increase in small-cells should be non-positive for both SPs. This can be analytically expressed via the two corresponding inequalities:
\begin{align}\label{Eqn:Boundary Point NE}
\nonumber &\lambda_S{R_S^0}^{-\alpha}-{R_M^0}^{-\alpha}-\frac{\alpha \lambda_S^2B_{i,S}^0R_0}{N_f}{R_S^0}^{-\alpha-1}+\\
&\frac{(B_i-B_{i,S}^0)R_0}{N_m}{R_M^0}^{-\alpha-1}
\le 0, \text{ for } i=1,2.
\end{align}
Here, $R_S^0$ and $R_M^0$ are defined as follows:
\begin{subequations}
\begin{align}
&R_S^0=\frac{\lambda_S(B_{1,S}^0+B_{2,S}^0)R_0}{N_f},\\
&R_M^0=\frac{(B_1-B_{1,S}^0+B_2-B_{2,S}^0)R_0}{N_m}.
\end{align}
\end{subequations}
%
%While this exact condition requires some calculation, the following corollary gives a simpler sufficient condition for this to be true.
%\begin{corollary}
%In subcase 2, if $\lambda_S\ge 2\alpha$, then the Nash equilibrium is always of the first type.
%\end{corollary}

{\it Case C: Only one constraint is violated.}
Without loss of generality, we assume at the Nash equilibrium without regulations, only SP 2's small-cell bandwidth allocation falls below the required threshold. In this case, the new Nash equilibrium is characterized by the following proposition:
\begin{proposition}\label{Prop:Both Violated NE}
In case C, the Nash equilibrium with regulatory constraints is one of the following two types:\\
Type I. Both SPs increase their small-cell bandwidth allocations to exactly the required amount, i.e, $B_{1,S}=B_{1,S}^0, B_{2,S}=B_{2,S}^0$.\\
Type II. Only SP 2 allocates exactly the required minimum amount of bandwidth to small-cells, i.e, $B_{1,S} > B_{1,S}^0, B_{2,S}=B_{2,S}^0$.
\end{proposition}

Note that equation (\ref{Eqn:Boundary Point NE}) also applies to give conditions when a type I equilibrium arises.

While the type I Nash equilibrium in both cases B and C indicate both SPs allocate exactly the required minimum amount to small-cells, they are quite different. In case B both SPs increase their small-cell bandwidth allocations, whereas in case C, one SP increases its small-cell bandwidth while the other SP decreases its small-cell bandwidth. Another difference that is worth pointing out is that in case C, the SP whose small-cell bandwidth allocation without regulations violates the constraint always operates at exactly the required minimum point at the new Nash equilibrium, while it will further increase its small-cell bandwidth beyond the minimum point in a type II equilibrium for case B.
\\

\begin{figure}[htbp]
\centering
\includegraphics[width=0.45\textwidth,height=0.35\textwidth]{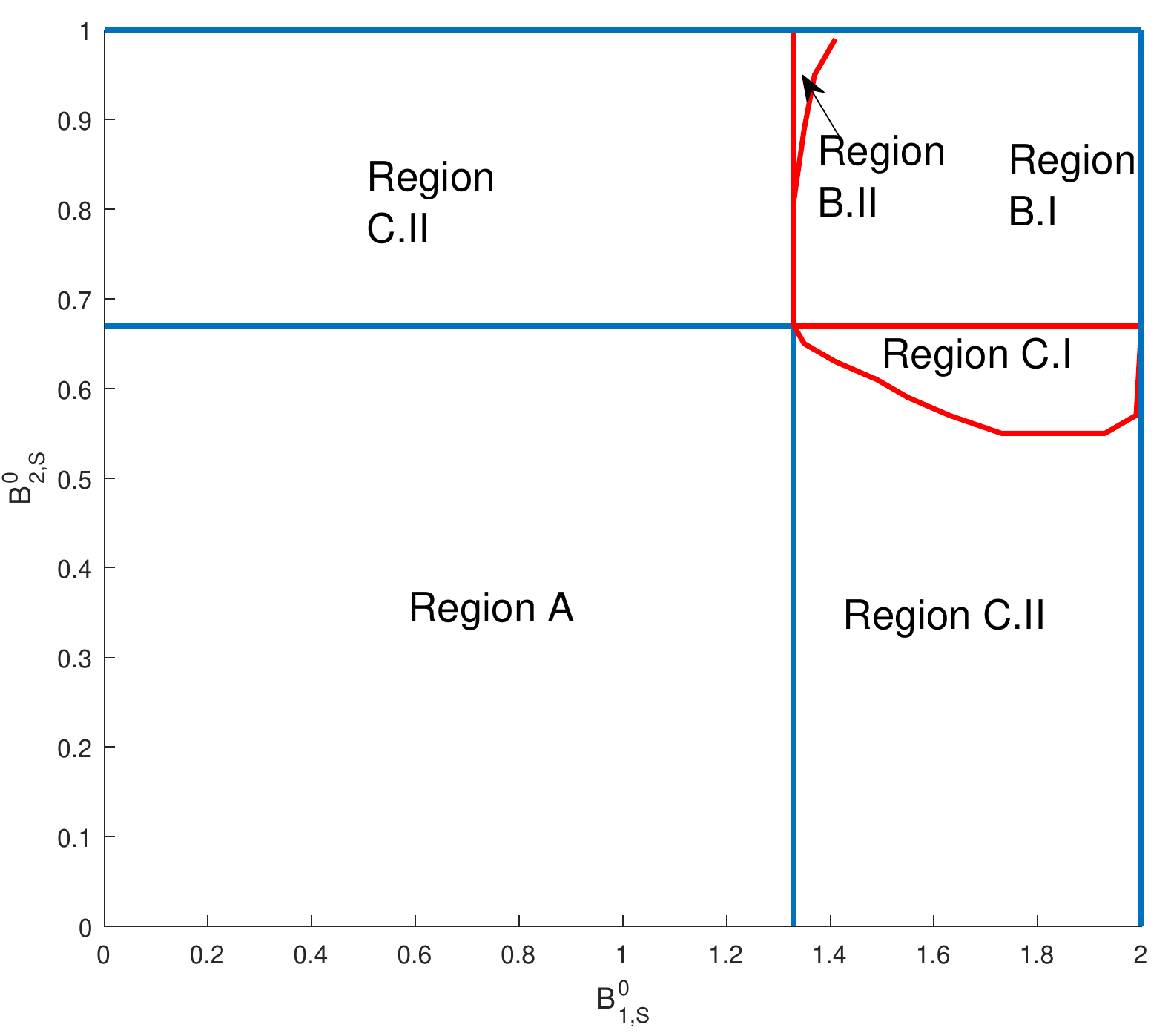}
\caption{Nash equilibrium regions for 2 SPs as the bandwidth regulations vary.}
\label{Fig:NE_Region}
\end{figure}

 Next we use a specific example in Figure \ref{Fig:NE_Region} to illustrate the different Nash equilibrium regions as a function of the small-cell bandwidth constraints discussed in the preceding cases. The system parameters for this case are: $\alpha=0.5, N_m=N_f=50, R_0=50, \lambda_S=2, B_1=2, B_2=1$. In this example the original equilibrium small-cell bandwidth allocations without the regulatory constraints are: $B_{1,S}=1.34, B_{2,S}=0.67$. Region A corresponds to the Nash equilibrium in case A, which is also the equilibrium without the regulatory constraints. Region B.I and Region B.II correspond to the type-I and type-II Nash equilibrium in case B where both constraints are violated at the original equilibrium, and the same rule applies to Region C.I and C.II.

\section{Social Welfare}\label{Sec:SW}
In this section we focus on social welfare analysis. In our previous work \cite{Opt7-Chen13}\cite{Competition5-Chen15}, we showed that for the set of $\alpha$-fair utility functions we use here, the bandwidth allocation at equilibrium is always socially optimal in both monopoly and competitive scenarios. With the additional regulatory constraints on the minimum amount of small-cell bandwidth allocations, this is not necessarily true. Obviously, if the equilibrium without regulations already satisfies the regulatory constraints, then the preceding result still holds, i.e., in case A in the previous section. Otherwise, a social welfare loss is incurred compared to the case without regulatory constraints. Denote $\text{SW}_{\text{wo}}^{*}, \text{SW}_{\text{w}}^{\text{NE}}$ as the equilibrium social welfare without and with regulatory constraints, respectively. The following theorem states that the loss in social welfare is lower bounded, and the worst point occurs at the scenario where the regulatory constraints require both SPs to allocate all bandwidth only to small-cells.

\begin{theorem}[Social Welfare] \label{Thm:SW}
Compared to the case without the regulatory constraints, social welfare loss is incurred when the following inequality is true:
\begin{equation}
\frac{N_f\lambda_S^{1/\alpha-1}}{N_f\lambda_S^{1/\alpha-1}+N_m} \sum\limits_{i\in \mathcal{N}}B_i< \sum\limits_{i\in \mathcal{N}}B_{i,S}^0.
\end{equation}
We have:
\begin{equation}
\frac{\text{SW}_{\text{w}}^{\text{NE}}}{\text{SW}_{\text{wo}}^{*}}
\ge \Big(\frac{N_f\lambda_S^{1/\alpha-1}}{N_f\lambda_S^{1/\alpha-1}+N_m}\Big)^{\alpha},
\end{equation}
where the bound is tight exactly when $B_{i,S}^0=B_i, \forall i\in \mathcal{N}$.
\end{theorem}

In practice, a spectrum regulator, such as the FCC, may seek to find an optimal way to allocate newly available spectrum so that the market equilibrium yields the largest social welfare. We next use our results to analyze the case where the spectrum regulator needs to allocate a total available new bandwidth $B$ to two competitive SPs. SP 1 and 2 each have initial licensed bandwidth $B_{1}^o$ and $B_{2}^o$, and get a proportion of the new bandwidth, denoted as $B_{1}^n$ and $B_2^n$. The initial bandwidth is free to use for either macro-cells or small-cells. In contrast, the new bandwidth can only be used for small-cells. As mentioned before this is motivated by the 3.5GHz band, where FCC regulates the power constraint to be very small, and therefore it can only be used for small-cell deployment \cite{FCC}.

The spectrum regulator needs to determine the optimal split of the new bandwidth such that the social welfare under market equilibrium is maximized. We consider the following three scenarios for any possible bandwidth partition $(B_1^n, B_2^n)$:

1) The optimal social welfare without regulatory constraints, $\text{SW}_{\text{wo}}^{*}$. Note, from \cite{Competition5-Chen15}, this is the same as the equilibrium social welfare without regulatory constraints.
This will be used as a benchmark.

2) The optimal social welfare with the regulatory constraints, which we denote as $\text{SW}_\text{w}^{*}$.

3) The equilibrium social welfare with regulatory constraints, $\text{SW}_{\text{w/}}^{\text{NE}}$.

The next theorem compares the three scenarios depending on the total amount of newly available bandwidth $B$.

\begin{theorem}\label{Thm:Three Scenarios}
Depending on the amount of new bandwidth $B$, there exists a bandwidth threshold $T$
\begin{equation}
T=\frac{(B_1^o+B_2^o)N_f\lambda_S^{1/\alpha-1}}{N_m},
\end{equation}
and we have the following conclusions:\\
1. If $B>T$, then  $\text{SW}_{\text{w}}^{\text{NE}}\le \text{SW}_{\text{w}}^{*} < \text{SW}_{\text{wo}}^{*}$. The first inequality is binding, i.e, $\text{SW}_{\text{w}}^{\text{NE}}= \text{SW}_{\text{w}}^{*} < \text{SW}_{\text{wo}}^{*}$, if and only if equation (\ref{Eqn:Boundary Point NE}) holds. \\
2. If $B\le \frac{(B_1^o+B_2^o)N_f\lambda_S^{1/\alpha-1}}{N_m}$, then  $\text{SW}_{\text{w}}^{\text{NE}}\le \text{SW}_{\text{w}}^{*} = \text{SW}_{\text{wo}}^{*}$. The first inequality is binding, i.e., $\text{SW}_{\text{w}}^{\text{NE}} = \text{SW}_{\text{w}}^{*} = \text{SW}_{\text{wo}}^{*}$, if and only if the following condition is met:
\begin{equation}
B_1^n\in \Big[  B-\frac{B_2^oN_f\lambda_S^{1/\alpha-1}}{N_m}, \frac{B_1^oN_f\lambda_S^{1/\alpha-1}}{N_m}  \Big], B_2^n=B-B_1^n.
\end{equation}
\end{theorem}

Theorem \ref{Thm:Three Scenarios} states that if the total amount of newly available bandwidth is too large, no matter if the two competing SPs maximize revenue or social welfare, we always have some social welfare loss compared to the case without regulatory constraints. This can be explained as follows. Using the set of $\alpha$-fair utility functions, without regulatory constraints the socially optimal bandwidth allocation strategy is to allocate bandwidth to macro- and small-cells based on a fixed proportion. If the total amount of newly available bandwidth is not large, simply following the original allocation satisfies the regulation requirement and is therefore socially optimal. However, when the amount of new bandwidth becomes large, since the new bandwidth is required to be allocated to small-cells only, the original optimal proportion would no longer be valid given the small-cell bandwidth constraints. As a result of this, social welfare loss relative to the original allocation scheme becomes inevitable. Further, note that the bandwidth threshold at which this loss begins occurring is proportional to $\frac{N_f}{N_m}$, so that when there are more fixed users willing to use small-cells, the threshold increases. It is also increasing in $\lambda_S$, the gain in spectral efficiency of small-cells and in the initial allotment of licensed bandwidth.

Theorem \ref{Thm:Three Scenarios} also indicates that when the amount of newly available bandwidth is below the threshold, there exists a bandwidth split that achieves the optimal benchmark social welfare. This result suggests that if a spectrum controller is planning to enforce bandwidth regulations on newly released bands, it should consider the possible impacts on the market equilibrium without regulations carefully. In particular, if the amount of newly available bandwidth is too large, imposing such regulations might lead to social welfare loss compared to the scenario where the regulations were not imposed. On the other hand, if the amount of new spectrum is small compared to the existing bands already licensed to SPs in the market, the influence on the market equilibrium from the introduction of bandwidth regulations on the new bands is minor and controllable, and therefore will not incur any loss in the social welfare.

\begin{figure}[htbp]
\centering
\includegraphics[width=0.45\textwidth,height=0.35\textwidth]{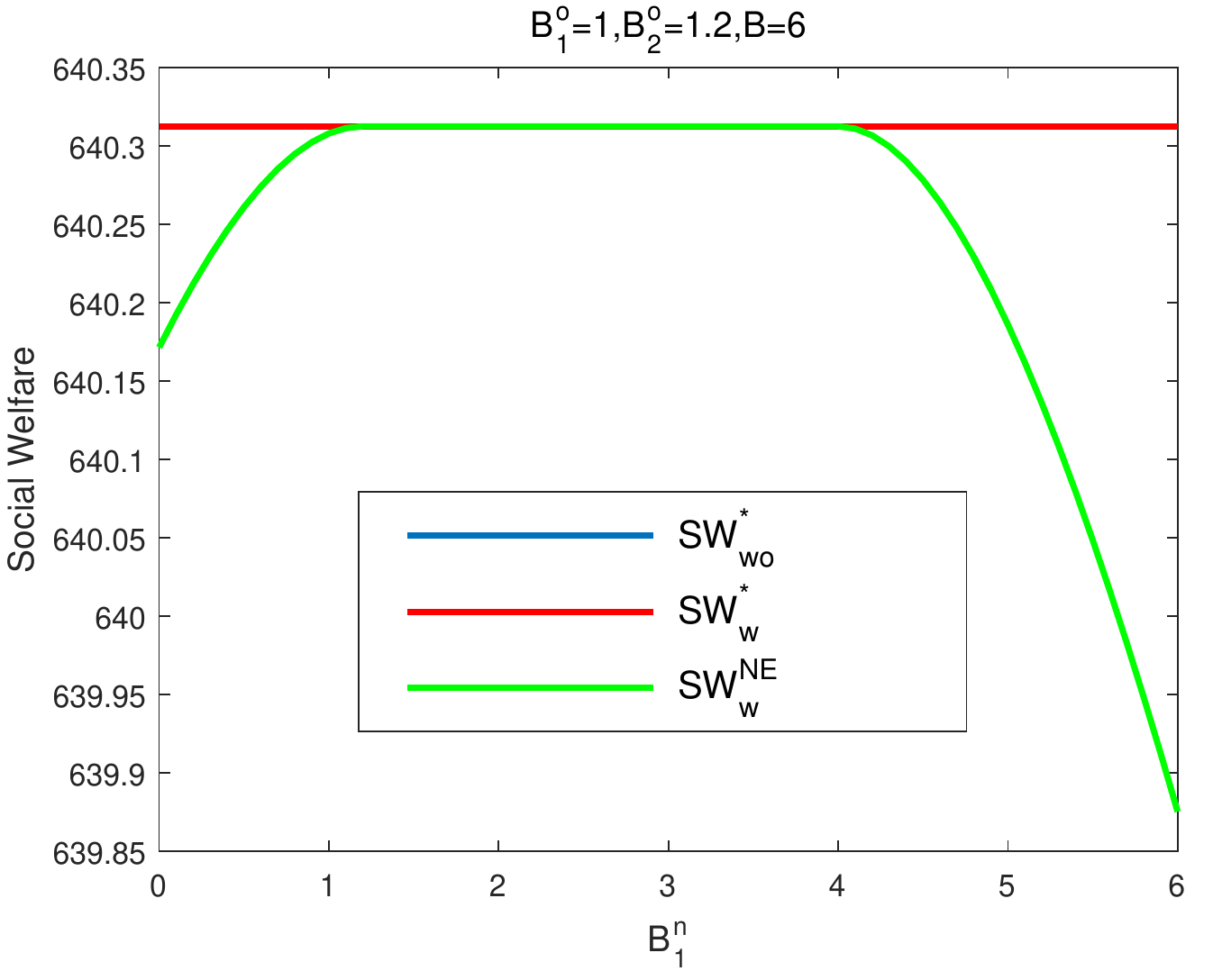}
\caption{Social welfare versus $B_1^n$ with small $B$. }
\label{Fig:SW_1}
\end{figure}

Figure \ref{Fig:SW_2} and \ref{Fig:SW_1} illustrate Theorem \ref{Thm:Three Scenarios}. The system parameters we use in both cases are: $\alpha=0.5, N_m=N_f=50, R_0=50,\lambda_S=4, B_1^o=1, B_2^o=1.2$. The Figures differ in the amount of new bandwidth. In Figure \ref{Fig:SW_2}, $B=10$, while in Figure \ref{Fig:SW_1}, $B=6$.  We can see that when the amount of newly available bandwidth is not large, there is a bandwidth split that achieves the optimal benchmark social welfare. However, when the amount of new bandwidth is large relative to the amount of original bandwidth of the SPs, there exists no possible bandwidth split schemes that achieve the optimal social welfare without the constraints.

%\begin{figure}[htbp]
%\centering
%\includegraphics[width=0.45\textwidth,height=0.35\textwidth]{SW_2.pdf}
%\caption{Social welfare versus $B_1^n$ with large $B$.}
%\label{Fig:SW_2}
%\end{figure}

\section{Conclusions}\label{Sec:Conclusions}
In this paper we considered the impact of bandwidth regulations on resource allocation in a HetNet. We showed that by imposing a required minimum bandwidth allocation to small-cells, the optimal bandwidth allocation strategies of SPs can change dramatically. While this change is relatively straightforward in the monopoly scenario, it turns out to be much more complicated in the competitive scenario with two SPs. Specifically, the existence and uniqueness of Nash equilibria are still preserved after adding the regulatory constraints. However, the equilibria can exhibit very different structures and characteristics as the constraints vary. We also showed that the introduction of such regulations may shift the equilibrium away from an efficient allocation, thus incurring some social welfare loss. Our results suggest that adding such regulatory constraints complicates the resource allocation schemes in the HetNets. While these constraints may be introduced by the spectrum regulator to address other concerns, they can ultimately reduce the social welfare achieved. For future directions, we are planning to study other policy considerations that we did not take into account in this paper, like the implications for innovation, spectrum caps, and the management of harmful interference.

%\clearpage

\clearpage

\begin{appendices}
\section{Proof of Theorem \ref{Thm:Monopoly Bandwidth Allocation}}
The proof of Theorem \ref{Thm:Monopoly Bandwidth Allocation} is straightforward and we can apply the results in \cite{Opt7-Chen13} directly. In particular, if the original optimal bandwidth allocation still holds with the added regulation constraints, then we are done. Otherwise we have to increase the small-cell bandwidth allocation. Both the revenue and social welfare are concave functions in the small-cell bandwidth allocation, and at the original equilibrium point the marginal revenue and social welfare increase with respect to per unit increase in bandwidth are equal for both macro- and small-cells. Hence, when the small-cell bandwidth increases, we enter the region where the marginal revenue and social welfare increase with respect to per unit increase in bandwidth for small-cells is smaller than that of macro-cells. As a result, the best option is to operate at the boundary point, i.e., allocating exactly the required minimum amount of bandwidth to small-cells.

\section{Proof of Theorem \ref{Thm:NE}}
As this is a concave game, to prove the existence and uniqueness of the Nash equilibrium, we can use the uniqueness theorem (Theorem 6) in Rosen's paper \cite{Rosen65}, which gives sufficient condition in terms of a certain matrix being negative definite. In our previous work \cite{Competition5-Chen15} it was proved that the required matrix is negative definite for the corresponding game without bandwidth restrictions. Here the only difference is that we have additional linear constraints on the bandwidth allocations, which does not have any effect on this result. Therefore, the same arguments also apply here.

As for the second part of the theorem, denote $R_S$ and $R_S^{\prime}$ as the average service rate in small-cells with and without the regulatory constraints, respectively. Suppose at the Nash equilibrium with constraints, the sum bandwidth allocation to small-cells is less than that without the regulatory constraints, then we have:
\begin{equation}
R_S^{\prime}<R_S.
\end{equation}

Denote $D_i=\frac{\partial S_i}{\partial B_{i,S}}$, it follows that:
\begin{align}
\nonumber  D_1+D_2=&\lambda_S\Big[2u^{\prime}(R_S)+R_Su^{\prime\prime}(R_S)\Big]-\\
&\Big[2u^{\prime}(R_M)+R_Mu^{\prime\prime}(R_M)\Big].
\end{align}

Since $2u^{\prime}(r)+ru^{\prime\prime}(r)$ decreases in $r$, and we know that at the Nash equilibrium without constraints, $D_1=D_2=0$, we can conclude that $D_1+D_2>0$ at the equilibrium with constraints. As a result, at least one of $D_1$ or $D_2$ must be greater than 0 at equilibrium. Without loss of generality, suppose $D_2>0$ at the equilibrium with constraints.

Given $D_2>0$ , it must be that $B_{2,S}^{\prime}=B_2$, and $D_1\le 0$. This is because if $D_1>0$ also holds, $B_{1,S}^{\prime}=B_1$ and it contradicts with the fact that $R_S^{\prime}<R_S$.

Then at the Nash equilibrium without constraints, we have:
\begin{align}
\nonumber D_1=&\lambda_Su^{\prime}(R_S)+\lambda_S^2\frac{B_{1,S}R_0}{N_f}u^{\prime\prime}(R_S)\\
&-u^{\prime}(R_M)-\frac{B_{1,M}R_0}{N_m}u^{\prime\prime}(R_M)\\
\nonumber =&\lambda_S\Big[u^{\prime}(R_S)+R_Su^{\prime\prime}(R_S)\Big]-\Big[u^{\prime}(R_M)+R_Mu^{\prime\prime}(R_M)\Big]\\
& -\lambda_S^2\frac{B_{2,S}R_0}{N_f}u^{\prime\prime}(R_S)+\frac{B_{2,M}R_0}{N_m}u^{\prime\prime}(R_M)=0. \label{Appen_B_Inequality1}
\end{align}

At the equilibrium with constraints, similarly we have:
\begin{align}\label{Appen_B_Inequality2}
\nonumber D_1=& \lambda_S\Big[u^{\prime}(R_S^{\prime})+R_S^{\prime}u^{\prime\prime}(R_S^{\prime})\Big]-
\Big[u^{\prime}(R_M^{\prime})+R_M^{\prime}u^{\prime\prime}(R_M^{\prime})\Big]\\
& -\lambda_S^2\frac{B_2R_0}{N_f}u^{\prime\prime}(R_S^{\prime})\le 0.
\end{align}

However, since $u^{\prime}(r)+ru^{\prime\prime}(r)$ decreases in $r$, $u^{\prime\prime}(r)<0$ and increases in $r$, and the fact that $R_S^{\prime}<R_S, R_M^{\prime}>R_M$, the inequality sign in (\ref{Appen_B_Inequality2}) should be reversed. Therefore we have a contradiction.

\section{Proof of Theorem \ref{Thm:SW}}
Applying the same arguments we used in proving Theorem \ref{Thm:Monopoly Bandwidth Allocation}, we know that since increasing the small-cell bandwidth allocation beyond the original equilibrium point only decreases the social welfare, then the worst case occurs at the point that all bandwidth is required to be allocated to small-cells.

\section{Proof of Theorem \ref{Thm:Three Scenarios}}
For scenario 2) and 3), as long as the sum of the small-cell bandwidth allocations of the two SPs at the equilibrium without the constraints is larger than the sum of the regulation constraints, then they are the same. This requires:
\begin{equation}
\frac{N_f\lambda_S^{1/\alpha-1}(B_1^o+B_1^n+B_2^o+B_2^n)}{N_f\lambda_S^{1/\alpha-1}+N_m}\ge B_1^n+B_2^n,
\end{equation}
which yields the following condition:
\begin{equation}
B\le \frac{(B_1^o+B_2^o)N_f\lambda_S^{1/\alpha-1}}{N_m}.
\end{equation}
Otherwise, if the preceding condition is not satisfied, the social welfare corresponding to the second scenario is also less than that corresponding to the first scenario, i.e, $\text{SW}_{\text{w}}^{*} < \text{SW}_{\text{wo}}^{*}$.

On the other hand, the only possible way for scenario 3) to achieve the optimal social welfare corresponding to scenario 1) is to ensure the Nash equilibrium is exactly the same as the one without the regulation constraints. This requires:
\begin{equation}
\frac{N_f\lambda_S^{1/\alpha-1}(B_1^o+B_1^n)}{N_f\lambda_S^{1/\alpha-1}+N_m}\ge B_1^n,
\frac{N_f\lambda_S^{1/\alpha-1}(B_2^o+B_2^n)}{N_f\lambda_S^{1/\alpha-1}+N_m}\ge B_2^n,
\end{equation}
which can be simplified to:

\begin{subequations}
\begin{align}
&B\le \frac{(B_1^o+B_2^o)N_f\lambda_S^{1/\alpha-1}}{N_m},\\
&B_1^n\in \Big[  B-\frac{B_2^oN_f\lambda_S^{1/\alpha-1}}{N_m}, \frac{B_1^oN_f\lambda_S^{1/\alpha-1}}{N_m}  \Big].
\end{align}
\end{subequations}

When $\text{SW}_{\text{w}}^{*} < \text{SW}_{\text{wo}}^{*}$, it means the required minimum sum bandwidth allocation to small-cells is larger than the sum bandwidth in small-cells at the equilibrium without constraints. Since we know that at the original equilibrium the social welfare is maximized and the social welfare is a concave function with respect to the sum bandwidth in small-cells, in this case the social welfare maximizing point with the constraints is therefore exactly the required minimum small-cell bandwidth point, i.e, when $B_{1,S}+B_{2,S}=B_{1,S}^0+B_{2,S}^0$. The only possibility for scenario 3) to achieve this is to ensure $B_{1,S}=B_{1,S}^0, B_{2,S}=B_{2,S}^0$ at the Nash equilibrium with constraints. As a result, equation (\ref{Eqn:Boundary Point NE}) becomes exactly the condition for $\text{SW}_{\text{w}}^{\text{NE}}=\text{SW}_{\text{w}}^{*} $.

\end{appendices}

\end{document}